# Scale-dependent permeability and formation factor in porous media: Applications from percolation theory

Misagh Esmaeilpour[1], Behzad Ghanbarian[1*], Feng Liang[2], and Hui-Hai Liu[2]

[1] Porous Media Research Lab, Department of Geology, Kansas State University, Manhattan 66506 KS, USA

[2] Aramco Services Company: Aramco Research Center—Houston, 16300 Park Row, Houston 77084 TX, USA

* Corresponding author's email address: ghanbarian@ksu.edu

**Abstract**

Understanding porous media properties and their scale dependence have been an active subject of research in the past several decades in hydrology, geosciences and petroleum engineering. The scale dependence of flow in porous media is attributed to small- and large-scale heterogeneities, such as pore size distribution, pore connectivity, long-range correlations, fractures and faults orientations, and spatial and temporal variations. The main objective of this study was to investigate how permeability ($k$) and formation factor ($F$) vary with sample dimension at small scales by means of a combination of pore-network modeling and percolation theory. For this purpose, the permeability and formation factor were simulated in twelve three-dimensional pore networks with different levels of pore-scale heterogeneities. Simulations were carried out at five different network sizes, i.e., 1130, 2250, 3380, 4510 and 6770 microns ($\mu m$). Four theoretical models were also developed based on percolation theory to estimate the scale dependence of permeability and formation factor from the pore-throat radius distribution. In addition, two other theoretical scale-dependent permeability models were proposed to estimate permeability at different scales from the pore-throat radius distribution and formation factor. Comparing theoretical estimations with numerical simulations showed that the proposed models estimate the scale dependence of permeability and formation factor reasonably. The calculated relative error (RE) ranged between -3.7 and 3.8% for the permeability and between 0.21 and 4.04% for the formation factor in the studied pore-networks.

# 1. Introduction

Modeling flow and transport in porous media has been an active subject of research in various disciplines, such as groundwater hydrology, petroleum and chemical engineering, soil physics, and geoscience. Since properties of porous media are measured at various scales, e.g., pore, core and reservoir/aquifer, understanding the effect of scale is essential, particularly for relating a property's value at a larger scale (e.g., reservoir) to its value at a smaller one (e.g., core). For this purpose, applying scaling techniques is necessary to transfer knowledge from one scale to another. This can happen by identifying governing mechanisms at smaller scales and then portraying their manifestation at larger scales [1].

The influence of measurement scale (or sample volume) on physical and hydraulic properties of porous media has been known for decades [2–7], and various scaling approaches have been proposed to study the scale dependence of flow and transport in porous media. One of the pioneer models is the Miller-Miller similar-media theory [8] in which all regions in a porous medium are assumed to be structurally identical magnifications of a reference location. This approach has been widely used to classify porous media based on their hydraulic properties, such as capillary pressure and hydraulic conductivity curves. Generally speaking, the Miller-Miller theory is valid as long as media are similar either with respect to their pore space characteristics or their hydraulic properties. Recently, Sadeghi et al. [9] discussed that similarity is not the only required condition. They demonstrated that the interrelation between the capillary pressure and hydraulic conductivity curves is also required for scaling purposes and classifying porous media.

Experimental measurements [10–12] and numerical simulations on rock images [13–15] indicate that permeability, $k$, increases with the increase in sample volume (or scale). However, beyond a critical volume, interpreted as the minimum scale of an equivalent homogeneous medium [16], $k$ remains approximately constant. The critical size or volume is known as the representative elementary volume (REV), the smallest sample size above which $k$ does not vary with size [17]. Schulze-Makuch et al. [10] conducted a comprehensive study of scale-dependent permeability by analyzing experimental measurements from 39 different media. They suggested a power-law scaling relationship to correlate the increase in $k$ with the sample volume $V_s$ as follows:

$$k = CV_s^m \qquad (1)$$

where the constant $C$ and scaling exponent $m$ characterizing the medium's heterogeneity are empirical. It is not known whether the scaling exponent $m$ in soils depends on the texture [11]. It has been noted that the structure of media affects the value of $m$. For example, Schulze-Makuch et al. [10] found $m = 0.51$ in heterogeneous fractured media and $0.55 < m < 0.83$ in double-porosity media. However, Fallico et al. [18] reported a substantially smaller value ($m = 0.029$) for a tank filled by sand with a high percentage (76%) of grains between 0.063 and 0.125 mm. Negative exponents, i.e., $m = -0.06$ and $-0.05$ were also reported in soils [19].

Ghanbarian et al. [20] applied a machine-learning method called the contrast pattern aided regression (CPXR) and proposed scale-dependent functions to estimate permeability from other porous media properties using samples from the UNSODA database. They showed that by including sample dimensions, i.e., sample internal diameter and height (or length), $k$ estimations were substantially improved. However, such functions and the power-law model given in Eq. (1) are purely empirical, and because of their empiricism the interpretation of the parameters (i.e., $C$ and $m$) and their variations from one soil/rock sample to another is not clear.

Hopmans et al. [21] stated that the inherent complexity of flow in heterogeneous media and the need to integrate theory with experiment demand innovative and multidisciplinary research efforts to overcome restrictions imposed by current understanding of scale dependence of flow and transport. For example, Hyun et al. [7] treated a rock as a truncated random fractal and studied the scale dependence of permeability using a stochastic scaling theory.

An explicit theoretical expression for the scale dependence of $k$ can be derived in the context of percolation theory [22,23]. Hunt [24] considered an anisotropic medium whose horizontal connectivity was greater than its vertical one and rescaled the medium's axes to have equal conductances in each direction. The transformed medium, accordingly, turned into an isotropic system with elongated volume. Hunt [24] then combined concepts from percolation theory with the power-law pore-throat size distribution and proposed the following theoretical relationship to characterize the scale dependence of permeability across scales:

$$k(L) = k_{REV}\left[1 - \left(1 - \left(\frac{r_{tmin}}{r_{tmax}}\right)^{3-D_p}\right)\left(\frac{l_{t0}}{l_{t0}+L}\right)^{\frac{1}{\nu}}\right]^{\frac{2}{3-D_p}} \quad (2)$$

where $D_p$ is the pore space fractal dimension characterizing the size distribution of pore throats, $k_{REV}$ is the REV value of permeability, $r_{tmin}$ and $r_{tmax}$ are the minimum and maximum pore-throat radii in the medium, $l_{t0}$ is the typical pore-throat length, $L$ is the system size, and $\nu$ is the correlation length scaling exponent whose universal value is 0.88 in three dimensions [25]. Hunt [24] set $r_{tmax}/r_{tmin} = 5000$, $D_p = 2.95$, $l_{t0} = 1$, and $k_{REV} = 8.2 \times 10^{-10}$ m$^2$ (REV hydraulic conductivity = 0.008 m/s), compared theoretical estimations from Eq. (2) with experimental data from Schulze-Makuch [26] collected from various sites within a carbonate-rock aquifer in southeastern Wisconsin and found generally well agreement.

More recently, Daigle [27] combined the scale dependence of percolation threshold with a permeability model based on the Katz and Thompson [28] approach and fractal properties of porous media. Similar approach was applied by Davudov and Moghanloo [29] to study the scale dependence of permeability in shales. However, both models assume that porous media are fractal, and their pore-throat size distributions follow the power-law probability density function.

## 2. Objectives

In the ingenious scale-dependent permeability model of Hunt [24], Eq. (2), the pore-throat size distribution was approximated by the power-law probability density function. However, lognormal [30,31], Weibull [32,33], or mixed Gaussian [34] distribution might be a more accurate representation in some porous media. There are also some rocks whose pore-throat size distributions do not conform to any type of probability density functions [35–39]. In addition, Eq. (2) scales down permeability using its REV value, while typically upscaling permeability is desired. In the Hunt [24] article, Eq. (2) was compared with experimental measurements whose pore space properties were not available. Therefore, the main objectives of this study are to: (1) generalize the Hunt [24] approach to be independent of the shape of pore-throat size distribution, (2) compare the proposed generalized model with individual pore networks whose pore structures are known, and (3) extend the percolation-based model to formation factor and its scale dependence in porous media.

## 3. Pore-network modeling

Pore-scale numerical simulations and pore-network modeling have been successfully used to study flow and transport in porous media [35,40–42]. In what follows, we first explain three-dimensional pore-networks generation and then describe flow simulations in such networks.

### 3.1. Generating pore networks

To investigate the effect of scale on permeability and formation factor in porous media with different levels of heterogeneity, three different pore-throat radius ranges, i.e., 0.1-10, 1-50 and 10-75 $\mu m$ were considered. Within each range, four networks were constructed using different values of the Weibull distribution parameters, as described in detail below. Overall, twelve pore networks were generated using the open-access code developed by Valvatne [43]. For this purpose, we generated cubic lattices with fixed coordination number $Z = 6$ and pore-throat length $l_t = 100$ $\mu m$. Each pore network was composed of cylindrical pore throats and spherical pore bodies.

The size distribution of pore throats conformed to the following truncated Weibull probability density function

$$r_t = (r_{tmax} - r_{tmin})(-\delta \ln(x(1 - e^{(1/\delta)}) + e^{(1/\delta)}))^{1/\gamma} + r_{tmin} \quad (3)$$

where $\delta$ and $\gamma$ are the Weibull distribution shape factors, $x$ is a randomly generated number between 0 and 1, $r_t$ is the pore-throat radius, and $r_{tmin}$ and $r_{tmax}$ are the smallest and largest pore-throat radii, respectively, in the network.

The pore-body radius was accordingly determined based on the following relationship [43]:

$$r_b = \max\left(\zeta \frac{\sum_{i=1}^{n} r_{ti}}{n}, \max(r_{ti})\right) \quad (4)$$

in which $n$ is the number of pore throats connected to the same pore body and $\zeta$ is an aspect ratio whose distribution follows the truncated Weibull probability density function. In this study, we set $\zeta = 0$ meaning that the pore-body radius has the same size as the largest connected pore throat.

To generate pore networks of various pore-scale heterogeneities, we used different values of the $r_{tmax}/r_{tmin}$ ratio and Weibull distribution shape factors $\delta$ and $\gamma$. The summary of properties of all the generated pore-networks is presented in Table 1. To study the scale dependence of permeability and formation factor, we used pore networks of sizes 1130, 2250, 3380, 4510 and 6770 $\mu m$, which indicate the length of each side of the pore-network cube.

### 3.2. Simulating flow in pore networks

Permeability and formation factor were simulated using the "poreflow" pore-scale simulator, also developed by Valvatne [43]. The value of permeability was determined using Darcy's law

$$k = \frac{\mu q_t L}{A_t(P_{inlet}-P_{outlet})} \qquad (5)$$

where $\mu$ is the fluid viscosity, $A_t$ is the medium cross-sectional area, $L$ is the length, $q_t$ is the total flow rate, and $P_{inlet} - P_{outlet}$ indicates a pressure difference between the inlet and outlet. The total flow rate was calculated by solving for pressure throughout the network under the steady state flow condition while mass conservation was taking place at each pore body as follows

$$\sum_j q_{ij} = 0 \qquad (6)$$

in which $i$ represents each of the pore bodies and $j$ denotes all the pore throats connecting to pore body $i$. For this equation to be in effect we should suppose that viscous pressure drops are negligible compared to capillary pressure. In Eq. (6), $q_{ij}$ is the flow rate between two pore bodies and depends on the hydraulic conductance $g_{hij}$, the distance between the centers of the two pore bodies $l_{ij}$, and the pressure difference $\Delta P_{ij}$ as follows

$$q_{ij} = \frac{g_{hij}}{l_{ij}} \Delta P_{ij} \qquad (7)$$

The fluid conductance between two pore bodies was determined using the harmonic mean of contributing conductances

$$\frac{l_{ij}}{g_{hij}} = \frac{l_{bi}}{g_{hi}} + \frac{l_t}{g_{ht}} + \frac{l_{bj}}{g_{hj}} \qquad (8)$$

where $l_{bi}$ and $l_{bj}$ indicate the distance in between the center of pore body to the interface where pore body and pore throat meet, and $l_t$ is the pore-throat length [43].

Under the laminar flow conditions, the hydraulic conductance of a pore with irregular cross section is given by

$$g_h = c \frac{A_p^2 G}{\mu} \qquad (9)$$

in which $c$ is a constant whose value is 0.6, 0.5623 and 0.5 for equilateral triangle, square, and circular pores, respectively, $G$ is the shape factor, and $A_p$ is the pore cross section.

Formation factor $F$ in the context of electrical flow is analogous to absolute permeability and hydraulic flow [43]. It is defined as the ratio of saturated medium resistivity, $R_o$, to brine resistivity, $R_w$ ($F = R_o/R_w$). $R_o$ can be determined from Ohm's law as follows:

$$R_o = \frac{A_t \Delta V}{a_t L} \qquad (10)$$

where $\Delta V$ is the voltage drop and $a_t$ is the total current flow. Accordingly, the electrical conductance $g_e$ is given by

$$g_e = \frac{A_w}{R_w} \qquad (11)$$

where $A_w$ is the cross-sectional area that is occupied by the brine in the pore. From Eq. (11) and Ohms' law we can write [43]

$$a_t = g_e \Delta V, \qquad (12)$$

then $R_o$ and eventually formation factor $F$ are computed.

**4. Theoretical modeling**

**4.1. Percolation theory**

Percolation theory from statistical physics provides a theoretic framework to study connections between macroscopic quantities and underlying microscopic properties in homogeneous and heterogeneous networks [44,45]. Although initial models were proposed based on bond and site percolation classes and regular lattices [25], more realistic and representative models were developed using irregular and disorder lattices [46,47] and the continuum percolation class [48,49]. In what follows, we apply concepts from percolation theory to generalize the methodology proposed first by Hunt [24] and establish a general relationship between the critical pore-throat radius and pore space characteristics, such as pore-throat radius distribution and typical pore-throat length as well as the system size.

The fractal nature of clusters in a percolating system underlies interesting scale-dependent transport modes e.g., permeability and formation factor. Within the percolation theory framework, correlation length gives a measure of the largest length scale at which non-Euclidean or fractal geometry effects are observed. In an infinite system, excluding the percolating (infinite) cluster, the mean distance between any two sites on the same finite cluster, known as the correlation length $\chi$, is given by [25,45]

$$\chi = \chi_0 (p - p_c)^{-\nu}, p > p_c \tag{13}$$

where $p$ is the fraction of bonds (or sites) that are occupied or present, $p_c$ is the percolation threshold, $\nu$ is the critical scaling exponent whose value is 0.88 in three dimensions [25], and $\chi_0$ is the typical bond length. For length scales smaller than the correlation length ($L < \chi$), the system is heterogeneous and statistically self-similar fractal, while for length scales larger than $\chi$ the system is macroscopically homogeneous and follows Euclidean geometry (Fig. 1).

Let us map a porous medium into a network of cylindrical pore tubes. To apply the concept of correlation length from percolation theory to porous media with irregular pore networks, one may rewrite Eq. (13) as follows

$$\chi = \chi_0 (f - f_c)^{-\nu} \tag{14}$$

in which $f = \frac{V}{V_t}$ is the volume fraction and $f_c = \frac{V_c}{V_t}$ is the critical volume fraction of pores. $V$ represents the pore volume, $V_c$ is the critical volume, and $V_t$ is the total volume of pores.

Rearranging Eq. (14) gives

$$f = f_c + \left(\frac{\chi_0}{\chi}\right)^{\frac{1}{\nu}} \tag{15}$$

Eq. (15) gives $f = f_c$ for a system of infinite size. However, for $\chi < \chi_0$, Eq. (15) returns $f > 1$, which is an unphysical limit. Following Hunt [24], Eq. (15) can be approximately corrected as follows:

$$f = f_c + \left(\frac{\chi_0}{\chi + \chi_0}\right)^{\frac{1}{\nu}} \tag{16}$$

Such a modification was successfully evaluated by Hunt [24] to compare scale-dependent permeability estimations with experiments and by Ghanbarian et al. [50] to estimate scale-dependent tortuosity in porous media.

Replacing the typical bond length $\chi_0$ and the correlation length $\chi$ respectively with the typical pore-throat length $l_{t0}$ and the system length $L$ in Eq. (16) gives

$$f = f_c + \left(\frac{l_{t0}}{L + l_{t0}}\right)^{\frac{1}{\nu}} \tag{17}$$

In percolation theory, the critical volume of pores $V_c$ can be determined by integrating $l_t r_t^2 f(r_t)$ between $r_{tc}$ and $r_{tmax}$ as follows [44]

$$V_c \propto \int_{r_{tc}}^{r_{tmax}} l_t r_t^2 f(r_t) dr_t \tag{18}$$

where $r_t$ is the pore-throat radius, $l_t$ is the pore-throat length, and $f(r_t)$ represents the pore-throat radius distribution.

Similarly, the total volume of pores is given by

$$V_t \propto \int_{r_{tmin}}^{r_{tmax}} l_t r_t^2 f(r_t) dr_t \tag{19}$$

Accordingly, the critical fractional pore volume can be determined from the ratio $V_c/V_t$

$$f_c = \frac{V_c}{V_t} = \frac{\int_{r_{tc}}^{r_{tmax}} l_t r_t^2 f(r_t) dr_t}{\int_{r_{tmin}}^{r_{tmax}} l_t r_t^2 f(r_t) dr_t} \tag{20}$$

Following Neuman and his coworkers [51,52], Hunt [24] argued that the axes of an anisotropic system can be rescaled to give equal conductances in each direction. Imagine a rectangular system with equal horizontal dimensions but a vertical dimension shorter than the horizontal ones. If the horizontal permeability is 100 times greater than the vertical permeability, the vertical dimension should be 10 times shorter than the horizontal dimensions to have equal conductances in all directions. In such a case, the correlation length would be greater than the horizontal dimension but smaller than the vertical dimension. However, for transport through a system to be truly 3D, all dimensions of the system should be greater than the correlation length $\chi$. This means that such a transformed medium would be quasi one-dimensional with percolation threshold near 1 [24,44].

Following Hunt [24], to determine the scale dependence of the critical pore-throat radius as the critical volume fraction approaches 1 (the fully 1D limit), we replace $f$ in Eq. (17) with $f_c$ from Eq. (20) and let $f_c \to 0$ in Eq. (17) to have

$$\frac{\int_{r_{tc}}^{r_{tmax}} l_t r_t^2 f(r_t) dr_t}{\int_{r_{tmin}}^{r_{tmax}} l_t r_t^2 f(r_t) dr_t} = \left(\frac{l_{t0}}{L + l_{t0}}\right)^{\frac{1}{\nu}} \tag{21}$$

Eq. (21) is the general relationship implicitly linking the critical pore-throat radius to the pore-throat size distribution, typical pore-throat length, and system size. Since $r_{tc}$ in Eq. (21) is not an explicit function of $L$, we numerically determine its value. Note that, following Hunt [24], the original $f_c$ in Eq. (17) was set equal to 0 for convenience and simplicity.

In the following, we propose different models based on the relationship between permeability and/or formation factor and the critical pore-throat radius to estimate their scale dependency.

**4.2. Estimating the scale dependence of permeability from pore-throat size distribution**

A long-standing problem in petroleum engineering and many other research disciplines has been estimating permeability at a larger scale from its value measured and/or determined at a smaller scale. To address the effect of scale on the permeability, we invoke the critical path analysis (CPA) approach and percolation theory. Katz and Thompson [28] were first to apply concepts from CPA to estimate permeability from formation factor and critical pore-throat radius in porous media. They proposed

$$k = \frac{r_{tc}^2}{C_{CPA} F} \tag{22}$$

where $C_{CPA}$ is a constant whose value depends on geometrical properties of the pore space [53]. We combine Eq. (21) with Eq. (22), propose two different scale-dependent permeability models, and compare theoretical estimations with pore-network simulations.

Following Hunt [24], we assume that the critical pore-throat radius varies with the scale via Eq. (21). We further presume that permeability is dominantly controlled by the critical pore-throat radius. Accordingly, we set $k(L) \propto r_{tc}^2(L)$ and normalize permeability using its value at the smallest scale $L_{min}$ (i.e., $L = 1130$ μm) as follows

$$k(L) = k(L_{min}) \left[\frac{r_{tc}(L)}{r_{tc}(L_{min})}\right]^2 \quad (23)$$

Recall that $r_{tc}(L)$ can be numerically computed by solving Eq. (21) given that $f(r_t)$, $l_{t0}$, and $L$ are known for the pore networks studied here. Following Katz and Thompson [28], we determine the value of $r_{tc}(L_{min})$ from the mode of the pore-throat size distribution (see Table 1).

The power-law relationship between $k$ and $r_{tc}$ may not follow the quadratic relationship given in Eq. (23). Reanalyzing experimental data reported by Katz and Thompson [28] revealed an exponent equal to 2.36 (results not shown). Ghanbarian et al. [54] also found an exponent less than 2 (i.e., 1.90). We accordingly generalize Eq. (23) to have

$$k(L) = k(L_{min}) \left[\frac{r_{tc}(L)}{r_{tc}(L_{min})}\right]^\alpha \quad (24)$$

where the value of $\alpha$ can be determined from the simulations by fitting a power law to the permeability values simulated at $L = 1130$ μm versus the critical pore-throat radius derived from the mode of the pore-throat radius distributions.

### 4.3. Estimating the scale dependence of formation factor from pore-throat size distribution

Formation factor $F$ is another important porous medium's property that has been widely investigated. However, the theoretical modeling of its scale dependence has remained as an open question in the literature. In the following, we propose two scale-dependent $F$ models using the CPA approach and percolation theory.

Following Ewing and Hunt [55], one may invoke concepts from CPA to establish a theoretical relationship between the formation factor and the critical pore-throat radius i.e., $F \propto$ $r_{tc}^{-1}$. Applying this relationship in combination with Eq. (17) provides the following scale-dependent $F$ model

$$F(L) = F(L_{min}) \left[ \frac{r_{tc}(L)}{r_{tc}(L_{min})} \right]^{-1} \tag{25}$$

Similar to the scale-dependent permeability, the relationship between $F$ and $r_{tc}$ may not conform to an inverse linear equation and Eq. (25). Reanalyzing experimental measurements reported by Katz and Thompson [28] revealed an exponent equal to -0.44 (results not shown). We, therefore, propose the following model for the scale dependence of the formation factor

$$F(L) = F(L_{min}) \left[ \frac{r_{tc}(L)}{r_{tc}(L_{min})} \right]^{-\beta} \tag{26}$$

in which the value of the exponent $\beta$ can be determined from the simulations and by directly fitting a power law to the formation factors plotted against their corresponding $r_{tc}$ values.

### 4.4. Estimating the scale dependence of permeability from formation factor and pore-throat size distribution

In the Hunt [24] model, it is assumed that the value of critical pore-throat radius varies with the scale. In this section, we propose two other models based on the Katz and Thompson [28] relationship, Eq. (22), to estimate $k(L)$ from either $F(L)$ or pore-throat radius distribution and formation factor.

If pore-throat radius distribution does not significantly change with scale, one may assume that permeability is dominantly controlled by formation factor and set $k(L) \propto 1/F(L)$. Accordingly, normalizing permeability using its value at the smallest scale gives

$$k(L) = k(L_{min}) \frac{F(L_{min})}{F(L)} \tag{27}$$

Eq. (27) provides a simple relationship to determine the scale dependence of permeability from permeability measured/simulated at the smallest scale, $k(L_{min})$, and the formation factor measured/simulated at both scales. Although such a linear proportionality is valid in pore-network simulations where pore-throat radius distributions do not change from one scale to another, it may not be held in real rocks [56].

If the value of $k(L_{min})$ is not available, one may estimate it via the Katz and Thompson [28] model, Eq. (22). Combining Eq. (27) with Eq. (22) yields

$$k(L) = \frac{r_{tc}^2(L_{min})}{C_{CPA}F(L)} \tag{28}$$

To estimate the scale dependence of permeability via Eq. (28), one needs the formation factor measured/simulated at that scale and the critical pore-throat radius at the smallest scale. The latter can be determined from the pore-throat radius distribution. In this study, we set $C_{CPA} = 32$, following Friedman and Seaton [57].

### 4.5. Models evaluation criteria

To evaluate the accuracy of the proposed scale-dependent models, the root mean square log-transformed error (RMSLE) and the relative error (RE) values were calculated as follows

$$RMSLE = \sqrt{\frac{1}{N}\sum_{i=1}^{N}[\log(x_{est}) - \log(x_{sim})]^2} \tag{29}$$

$$RE = \frac{x_{est} - x_{sim}}{x_{sim}} \times 100 \tag{30}$$

where $N$ is the number of samples, and $x_{est}$ and $x_{sim}$ are, respectively, the estimated and simulated values.

### 5. Results

In this section, we present the results of comparing the pore-scale numerical simulations with the proposed theoretical models from percolation theory. Fig. 2 shows the pore-throat radius distributions for all the twelve pore networks studied here. As can be seen from Fig. 2 (also indicated in Table 1), Networks 1, 2, and 3 have three distinct ranges of pore-throat radius to study the scale dependence of permeability and formation factor in porous media of different levels of heterogeneity. Network 1 has the narrowest pore-throat radius distribution among all the networks (Fig. 2) and represents the most homogeneous medium, while Networks 2 and 3 represent respectively the intermediate and the most heterogeneous media in this study. We should also point out that as the parameter $\gamma$ in the Weibull distribution, Eq. (3), increases (Table 1), the average value of $r_t$ increases as well. However, the pore-throat radius distribution becomes narrower (Fig. 2). Accordingly, the level of heterogeneity decreases from Networks 1.1, 2.1, and 3.1 to Networks 1.4, 2.4, and 3.4.

The pore-body radius distributions of all the networks are presented in Fig. 3. Since we set $\zeta = 0$ in Eq. (4), the pore-body radius has the same size as the largest connected pore throat. As a result, the pore-body radius distributions of the networks resembled probability density functions similar to the pore-throat radius distributions.

To determine the exponents $\alpha$ in Eq. (24) and $\beta$ in Eq. (26), we plotted the permeability and formation factor values simulated at the smallest network with $L = 1130$ $\mu m$ versus the critical pore-throat radius and fitted the power-law function to the data. Results shown in Fig. 4 indicate $\alpha = 3.03$ and $\beta = -1.37$ with $R^2 > 0.94$. We reanalyzed the experimental data from Katz and Thompson [58] and numerical simulations from Berg [59] and found $\alpha = 2.34$ and $4.71$ and $\beta = -0.44$ and $-2.66$, respectively (results not shown). Our $\alpha = 3.03$ and $\beta = -1.37$ are in accord with the range obtained from the literature. $\beta = -1.37$ in Eq. (26), is not greatly different from the exponent -1 in Eq. (25). Accordingly, the scale-dependent formation factor estimations by these two equations should not be substantially different, as we show in what follows. For permeability, $\alpha = 3.03$ in Eq. (24), however, is greater, by a factor of 1.5, than the exponent 2 in Eq. (23).

## 5.1. Estimating the scale dependency of permeability and formation from pore-throat radius distribution

**- Network 1**

Fig. 5 presents the results of pore-scale numerical simulations of $k$ and $F$ and the theoretical estimations by the proposed models, Eqs. (23)-(26). As reported in Table 1, the value of $r_{tc}$ increases from Network 1.1 to 1.4. Therefore, based on the CPA and Eq. (22) one should expect the value of permeability to increase from Network 1.1 to 1.4 as well, as shown in Fig. 5. Using the same terminology and given that the relationship between $F$ and $r_{tc}$ is inverse, one should expect the value of formation factor to decrease from Network 1.1 to 1.4.

Fig. 5 also shows that the simulated permeability increases with increase in the network size, although the pore-throat radius distribution does not statistically vary with the scale. The theoretical estimations of the scale-dependent permeability are also presented in Fig. 5, and the corresponding RMSLE value for each model is also reported. For the permeability, the RMSLE value ranged from 0.0075 to 0.0187, and 0.0032 to 0.0156 for the estimations based on Eq. (23), and Eq. (24), respectively. For the formation factor, we found $0.0141 < \text{RMSLE} < 0.0194$ for Eq. (25) and $0.0112 < \text{RMSLE} < 0.0185$ for Eq. (26), as reported in Fig. 5.

We also report the values of relative error (RE) and relative absolute error (RAE) as well as their averages in Table 2. Generally speaking, Eq. (23) and Eq. (24) underestimated the scale-dependent permeability. This is confirmed via the negative RE values reported in Table 2.

Results from Network 1 show that Eq. (24) with average RE = -1.67% estimated $k(L)$ more accurately than Eq. (23) with average RE = -2.85%. For the formation factor, both models overestimated the scale-dependent $F$ in the studied networks. However, Eq. (26) with average RE = 3.24% estimated $F(L)$ slightly more precisely than Eq. (25) with average RE = 3.6% (Table 2).

Comparing the scale-dependent permeability and formation factor estimations shown in Fig. 5 indicate that overall the scale dependence of permeability was more precisely estimated than that of formation factor. This can be due to the fact that the hydraulic conductance of a cylindrical pore is proportional to its radius to the fourth power, while the electrical conductance proportional to the second power. As the exponent increases, flow is increasingly concentrated in fewer pathways, which become increasingly tortuous. This means that permeability is affected by the pore space structure and its heterogeneity more than formation factor.

- **Network 2**

Results from Network 2 are shown in Fig. 6. Similar to Network 1, permeability increases, while formation factor decreases with increase in the network size. Compared to Network 1, the value of permeability is nearly two orders of magnitude greater and the value of formation factor is one order of magnitude smaller in Network 2. Based on the permeability plots shown in Fig. 6, it seems that Network 2 has a greater REV value than Network 1. Although the permeability in Network 1 does not vary with the scale at network size of 6770 $\mu m$ (Fig. 5), its value tends to keep increasing with the scale in Network 2. This clearly shows that Network 2 is more heterogeneous than Network 1. This pore-scale heterogeneity is because the pore-throat radius distribution in the former is broader than that in the latter (Fig. 2).

As can be seen in Fig. 6, the proposed theoretical models estimated the scale-dependent permeability in Network 2 reasonably well. The value of RMSLE ranged from 0.0022 to 0.0103 for the estimations by Eq. (23) and from 0.0014 to 0.0077 for those by Eq. (24). We found the average RE value for Eqs. (23) and (24) equal to -1.42 and -0.42%, respectively. These values are less than those reported for Network 1 (see Table 2). This is most probably because Eqs. (23)-(26) were developed based on concepts from CPA, a theory that works best in heterogeneous media with broad pore-throat radius distributions [22,44]. The average RAE values reported for Eqs. (23) and (24) also indicate that Eq. (24) estimated the scale dependence of permeability more accurately than Eq. (23).

Similar to the results from Network 1, both Eqs. (25) and (26) overestimated the scale dependence of formation factor in Network 2. We found that, on average, Eq. (25) estimated $F(L)$ with RE = 2.14% and Eq. (26) with RE = 1.84%. Since both models overestimated $F$, the values of RE and RAE are the same (Table 2). Eq. (26) estimates $F(L)$ more precisely than Eq. (25) because the value of the exponent $\beta$ in Eq. (26) was optimized from the simulations at the smallest network size (i.e., 1130 $\mu m$).

**- Network 3**

The pore-throat radius distributions in Network 3 is broader than those in Networks 1 and 2 and, thus, it is the most heterogeneous network among these three cases. The increasing trend in the permeability at larger network sizes (e.g., 6770 $\mu m$) indicates that the REV has not reached (Fig. 7), which is similar to the results from Network 2 (Fig. 6). As reported in Fig. 7, we found 0.0022 < RMSLE < 0.0065 for Eq. (23) and 0.0013 < RMSLE < 0.014 for Eq. (24) in the estimation of the scale dependence of permeability in Network 3. Eq. (24) estimated $k(L)$ more accurately than Eq. (23) in Networks 3.3 and 3.4. However, the source of error in Networks 3.1 and 3.2 is not clear yet.

Based on the average RE values reported in Table 2, overall Eq. (23) estimated the scale dependence of permeability in Network 3 more precisely than Eq. (24). More specifically, we found RE = -0.53 and 0.7% and RAE = 0.9 and 1.15%, respectively, for Eqs. (23) and (24). Although Eq. (24) estimated the scale dependence of permeability more accurately than Eq. (23) in Networks 1 and 2, Eq. (23) has higher accuracy than Eq. (24) in Network 3. The reason is yet not clear and requires further investigations.

Results of the formation factor and its scale-dependent estimations are also presented in Fig. 7. Similar to Networks 1 and 2, Eq. (26) provided more accurate estimations of $F(L)$ than Eq. (25) (see the RMSLE values reported in Fig. 7). The average RE and RAE values reported in Table 2 confirmed the obtained results.

**5.2. Estimating the scale dependency of permeability from formation factor and pore-throat radius distribution**

In this section, we present the results of $k(L)$ estimated by Eq. (27) from $k(L_{\min})$ and $F(L)$ as well as by Eq. (28) from the pore-throat radius distribution and the simulated formation factor. Fig. 8 shows the results of $k(L)$ estimations via Eqs. (27) and (28). The RMSLE values for Network 1 ranged from 0.0020 to 0.0024 and from 0.0106 to 0.0163 for Eqs. (27) and (28), respectively. As can be seen in Table 3, for Eq. (27) and Network 1 we found the average RE = -0.38%, which is considerably less than those reported for Eqs. (23) and (24) in Table 2. For Eq. (28) and Network 1 the average RE value is 2.84%, greater than that for Eq. (27). The source of uncertainty in the estimations by Eq. (28) is most probably due to error in the estimation of $k(L_{min})$. Recall that Eqs. (23), (24), and (27) estimate $k(L)$ from $k(L_{\min})$ as well as the pore-throat radius distribution and/or formation factor, while in Eq. (28) $k(L_{min})$ is estimated from the pore-throat radius distribution and $F(L_{min})$.

Results of the scale-dependent permeability estimations using Eqs. (27) and (28) for Network 2 are also shown in Fig. 8. We found 0.0005 < RMSLE < 0.0007 for Eq. (27) and 0.0013< RMSLE < 0.0117 for Eq. (28). Similar to the results from Network 1, Eq. (27) provided more accurate estimations of $k(L)$ than Eq. (28). This is confirmed through the average RE and RAE values reported in Table 3. Comparing the RE values obtained from Eqs. (27) and (28) indicates that the error in the $k(L)$ estimation decreased from Network 1 to Network 2. As stated earlier, Network 2 is more heterogeneous, and its pore-throat radius distribution is broader compared to Network 1 (see Fig. 2). Since both Eqs. (27) and (28) are based on the CPA, one should expect the improvement of $k(L)$ estimations as the pore-throat radius distribution becomes broader.

The scale-dependent permeability estimations using Eq. (27) and Eq. (28) for Network 3 showed results similar to Networks 1 and 2. The RMSLE values for Network 3 ranged between 0.0003 and 0.0006 for Eq. (27) and between 0.0008 and 0.006 for Eq. (28). In Network 3, Eq. (28) underestimated $k(L)$ in all pore-networks except Network 3.1 (see negative RE values in Table 3). We found that the average RAE= 0.09 and 0.91% (Table 3), which clearly demonstrate that Eq. (27) provided more accurate estimations compared to Eq. (28).

## 6. Discussion

### 6.1. Models Accuracy

The RE and RAE values as well as their averages over all the networks for Eqs. (23)-(28) are reported in Tables 2 and 3. As can be seen, the RE value was found to be less than 5%, which indicates all the proposed models estimated the scale dependence of permeability and formation factor with high accuracy. Comparing the overall average RE and RAE values reported in Tables 2 and 3 demonstrates that Eq. (27) estimated $k(L)$ more accurately than other models developed in this study. More specifically, in the estimation of $k(L)$ we found the average RE = -1.6, -0.46, -0.19, and 1.2% for Eqs. (23), (24), (27), and (28), respectively. In fact, although Eq. (27) provides a simple relationship, it yielded the most accurate estimations among all the models for the networks studied here. After Eq. (27), Eq. (24) provided the most accurate estimations of $k(L)$.

For the formation factor, our results showed that Eq. (26) with the average RE = 2.04% estimated $F(L)$ slightly better than Eq. (25) with the average RE = 2.38%. In the estimation of the scale dependence of the formation factor the average RE and RAE values decreased from Network 1 to 3 (Table 2). In the estimation of $k(L)$, the same trend was observed for Networks 1 and 2. However, from Network 2 to 3, the average RE and RAE increased for Eq. (24). Further investigations are required to evaluate the proposed models using a broader range of pore networks with broader levels of heterogeneity.

**6.2. Limitations**

The proposed theoretical scale-dependent models for the permeability and formation factor were developed based on several fundamental assumptions one of which is that the pore-throat radius distribution does not greatly vary from one scale (or sample volume) to another. This presumption may be valid in pore-network simulations in which pore-throat radius distributions from different scales are statistically the same. However, it is not necessarily observed in natural porous media such as rocks that are heterogeneous in terms of pore space across scales [12,60,61].

Our proposed models are limited to experiments and/or simulations showing increasing trend in the permeability and decreasing trend in the formation factor with scale increase. That is because, based on Eq. (21), as the system size $L$ increases, the critical pore-throat radius $r_{tc}$ increases as well and approaches $r_{tmax}$ for $L \to \infty$. Since permeability is directly proportional and formation factor is inversely proportional to $r_{tc}$, $k$ is expected to increase and $F$ is expected to decrease as the system length increases according to Eqs. (23) to (26). Although the increasing trend in $k(L)$ has been widely observed in experiments [10–12,62] and numerical simulations on 3D images [13–15,63], there exist evidence in the literature [64–66] that $k$ may decrease as scale increases in certain cases.

In fact, the value of permeability depends on pore space characteristics, such as porosity, pore connectivity (coordination number), surface area, etc. If such properties vary across scales, depending on their trend and overall interactions, permeability may increase or decrease with increase in system size. Cui et al. [62] studied the scale dependency of permeability in shales of particles of sizes from nearly 0.2 to 20 mm (see their Fig. 10). They stated that shales crushed into particles of millimeter scale showed strong dual-pore structures. Cui et al. [62] argued that permeability in smaller particles most probably represents intact matrix properties in fractured reservoir rocks. However, at larger field scales fractures on scales from micrometers to meters may contribute to flow, and fractures and their networks have different transport properties than pores in the intact matrix. In another study, Tinni et al. [12] measured GRI permeability on crushed shale samples with particle sizes ranged from 0.7 to 6 mm. They reported permeability increase with particle size increase. Tinni et al. [12] argued that such an increase in GRI permeability was due to change in pore structure from one particle size to another and supported their statement using mercury porosimetry measurements.

## 7. Conclusion

Modeling the scale dependency of transport modes in porous media have been an active challenge in various research areas e.g., hydrology, geosciences and petroleum engineering. The scale dependence of flow and transport is attributed to small- and large-scale heterogeneities, such as pores and their size distribution, pore connectivity, long-range correlations, fractures and faults orientations, and spatial and temporal variations. In this study, we investigated the effect of scale on permeability and formation factor via pore-scale numerical simulations. Based on percolation theory and by extending the Hunt [24] approach, scale-dependent models were developed for permeability and formation factor. Comparing with pore-network simulations showed that all the proposed theoretical models estimated $k$ and $F$ accurately with relative errors less than 5%. Further investigations are required to evaluate the proposed models using a broader range of pore networks with broader levels of heterogeneity as well as experimental measurements.

**Acknowledgement**

ME is grateful to the College of Arts and Sciences as well as Graduate Students Council at Kansas State University for Travel Awards. BG acknowledges Kansas State University for supports through faculty startup funds.

Table 1. Salient properties of the twelve pore networks constructed in this study.

| Network | $r_b$ ($\mu m$) | $r_t$ ($\mu m$) | $\gamma$ | $\delta$ | $l_t$ ($\mu m$) | $Z$ | $r_{tc}$ ($\mu m$) | $\phi$ (%) |
|---|---|---|---|---|---|---|---|---|
| 1.1 | 0.1-10 | 0.1-10 | 12 | 0.2 | 100 | 6 | 8.7 | 4.3 |
| 1.2 | 0.1-10 | 0.1-10 | 18 | 0.2 | 100 | 6 | 9.1 | 4.7 |
| 1.3 | 0.1-10 | 0.1-10 | 24 | 0.2 | 100 | 6 | 9.35 | 4.9 |
| 1.4 | 0.1-10 | 0.1-10 | 30 | 0.2 | 100 | 6 | 9.5 | 5.1 |
| 2.1 | 1-50 | 1-50 | 12 | 0.2 | 100 | 6 | 44.2 | 32.1 |
| 2.2 | 1-50 | 1-50 | 18 | 0.2 | 100 | 6 | 46 | 34.2 |
| 2.3 | 1-50 | 1-50 | 24 | 0.2 | 100 | 6 | 47 | 35.4 |
| 2.4 | 1-50 | 1-50 | 30 | 0.2 | 100 | 6 | 47.5 | 36.1 |
| 3.1 | 10-75 | 10-75 | 12 | 0.2 | 100 | 6 | 67.6 | 43.9 |
| 3.2 | 10-75 | 10-75 | 18 | 0.2 | 100 | 6 | 69.7 | 46 |
| 3.3 | 10-75 | 10-75 | 24 | 0.2 | 100 | 6 | 70.7 | 47.1 |
| 3.4 | 10-75 | 10-75 | 30 | 0.2 | 100 | 6 | 71.5 | 47.8 |

[*] $r_b$ is pore-body radius, $r_t$ is pore-throat radius, $\gamma$ and $\delta$ are Weibull distribution parameters, $l_t$ is pore-throat length, $Z$ is pore coordination number, $r_{tc}$ is critical pore-throat radius, and $\phi$ is porosity.

Table 2. Calculated values of relative error (RE) and relative absolute error (RAE) for theoretical models developed to estimate the scale dependence of permeability and formation factor in twelve pore networks using Eq. (23), Eq. (24), Eq. (25), and Eq. (26).

| Pore network | Permeability $k$ | | | | Formation factor $F$ | | | |
| --- | --- | --- | --- | --- | --- | --- | --- | --- |
| | Eq. (23) | | Eq. (24) | | Eq. (25) | | Eq. (26) | |
| | RE | RAE | RE | RAE | RE | RAE | RE | RAE |
| 1.1 | -1.45 | 1.45 | 0.61 | 0.61 | 2.92 | 2.92 | 2.31 | 2.31 |
| 1.2 | -2.95 | 2.95 | -1.74 | 1.74 | 3.62 | 3.62 | 3.25 | 3.25 |
| 1.3 | -3.29 | 3.29 | -2.43 | 2.43 | 3.81 | 3.81 | 3.54 | 3.54 |
| 1.4 | -3.70 | 3.70 | -3.10 | 3.10 | 4.04 | 4.04 | 3.85 | 3.85 |
| Average | -2.85 | 2.85 | -1.67 | 1.97 | 3.60 | 3.60 | 3.24 | 3.24 |
| 2.1 | -0.45 | 0.45 | 1.26 | 1.26 | 1.68 | 1.68 | 1.19 | 1.19 |
| 2.2 | -1.36 | 1.36 | -0.28 | 0.28 | 2.10 | 2.10 | 1.78 | 1.78 |
| 2.3 | -1.81 | 1.81 | -1.10 | 1.10 | 2.32 | 2.32 | 2.11 | 2.11 |
| 2.4 | -2.07 | 2.07 | -1.54 | 1.54 | 2.44 | 2.44 | 2.29 | 2.29 |
| Average | -1.42 | 1.42 | -0.42 | 1.05 | 2.14 | 2.14 | 1.84 | 1.84 |
| 3.1 | 0.74 | 0.74 | 2.86 | 2.86 | 0.80 | 0.80 | 0.21 | 0.21 |
| 3.2 | -0.46 | 0.46 | 0.84 | 0.84 | 1.36 | 1.36 | 0.98 | 0.98 |
| 3.3 | -1.11 | 1.11 | -0.27 | 0.27 | 1.68 | 1.68 | 1.43 | 1.43 |
| 3.4 | -1.30 | 1.30 | -0.64 | 0.64 | 1.77 | 1.77 | 1.57 | 1.57 |
| Average | -0.53 | 0.90 | 0.70 | 1.15 | 1.40 | 1.40 | 1.05 | 1.05 |
| Overall average | -1.60 | 1.72 | -0.46 | 1.39 | 2.38 | 2.38 | 2.04 | 2.04 |

*values are in percentage

Table 3. Calculated values of relative error (RE) and relative absolute error (RAE) for theoretical models developed to estimate the scale dependence of permeability in twelve pore networks using Eq. (27) and Eq. (28).

| Pore network | Permeability $k$ | | | |
| --- | --- | --- | --- | --- |
| | Eq. (27) | | Eq. (28) | |
| | RE | RAE | RE | RAE |
| 1.1 | -0.42 | 0.42 | 3.80 | 3.80 |
| 1.2 | -0.42 | 0.42 | 2.45 | 2.45 |
| 1.3 | -0.33 | 0.33 | 2.57 | 2.57 |
| 1.4 | -0.34 | 0.34 | 2.53 | 2.53 |
| Average | -0.38 | 0.38 | 2.84 | 2.84 |
| 2.1 | -0.10 | 0.10 | 2.72 | 2.72 |
| 2.2 | -0.13 | 0.13 | 1.27 | 1.27 |
| 2.3 | -0.10 | 0.10 | 0.93 | 0.93 |
| 2.4 | -0.10 | 0.10 | 0.30 | 0.30 |
| Average | -0.11 | 0.11 | 1.31 | 1.31 |
| 3.1 | -0.07 | 0.07 | 1.38 | 1.38 |
| 3.2 | -0.11 | 0.11 | -0.14 | 0.17 |
| 3.3 | -0.11 | 0.11 | -0.68 | 1.06 |
| 3.4 | -0.08 | 0.08 | -0.65 | 1.03 |
| Average | -0.09 | 0.09 | -0.02 | 0.91 |
| Overall Average | -0.19 | 0.19 | 1.20 | 1.69 |

*values are in percentage

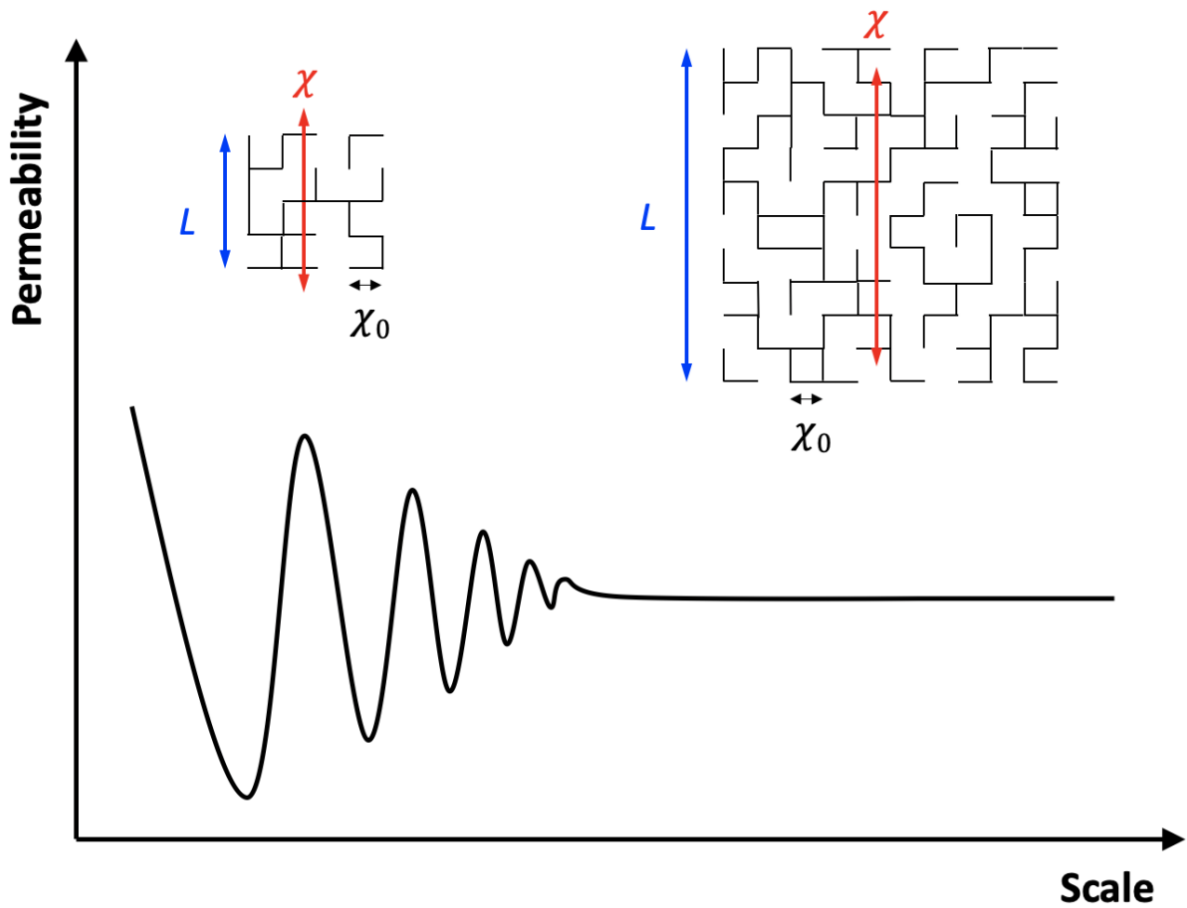

Fig. 1. The *schematic* plot of the scale dependence of permeability as well as the representative elementary volume (REV), the smallest size above which permeability does not vary with length. The correlation length, $\chi$, provides a measure of the largest length scale above which the system is macroscopically homogeneous, and the geometry is Euclidean ($L > \chi$). However, when the system size $L$ is less than the correlation length ($L < \chi$), the system is heterogeneous and statistically self-similar fractal. For transport through a system to be truly 3D, all dimensions of the system should be greater than the correlation length $\chi$.

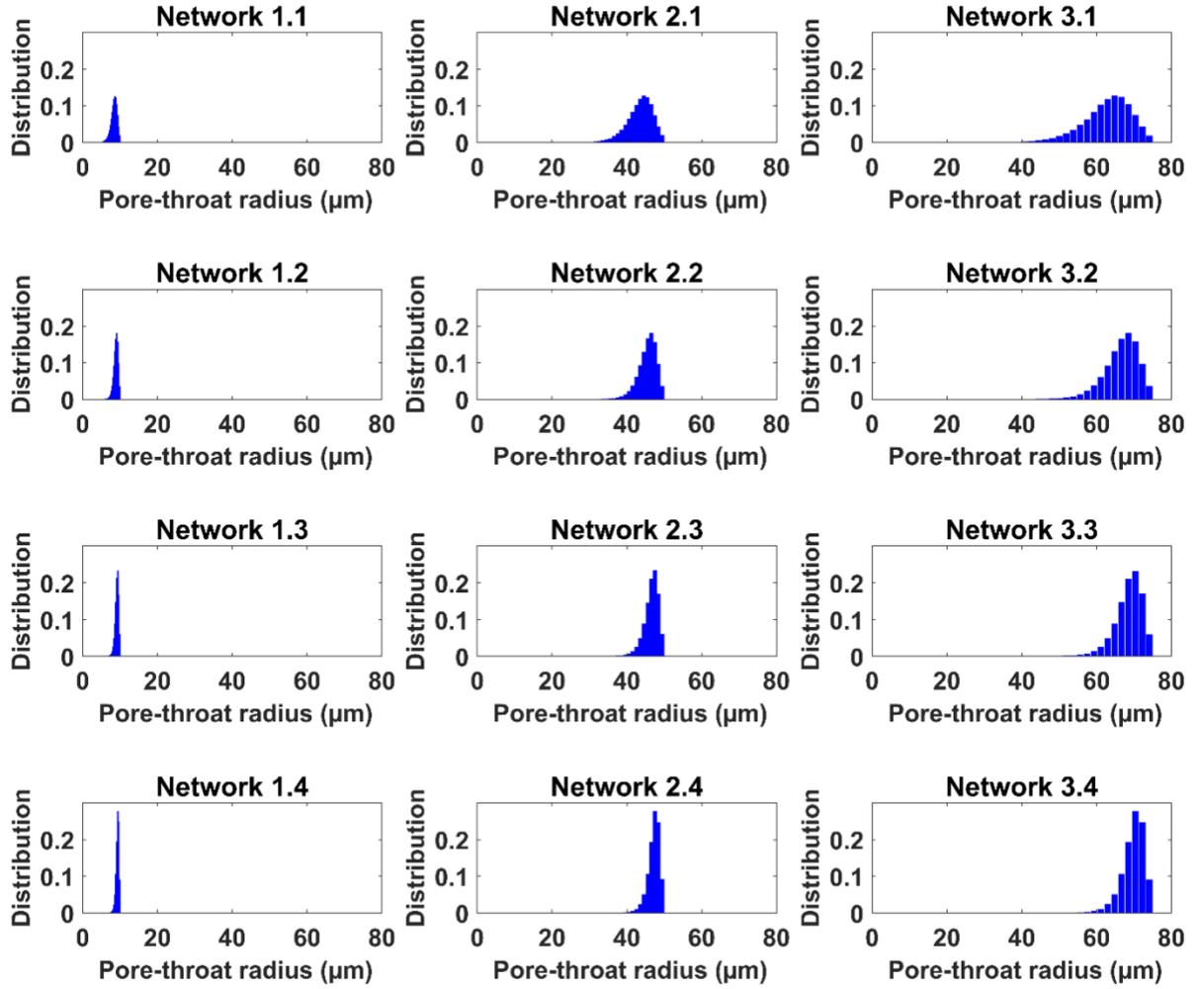

Fig. 2. Pore-throat radius distributions for twelve pore networks constructed in this study.

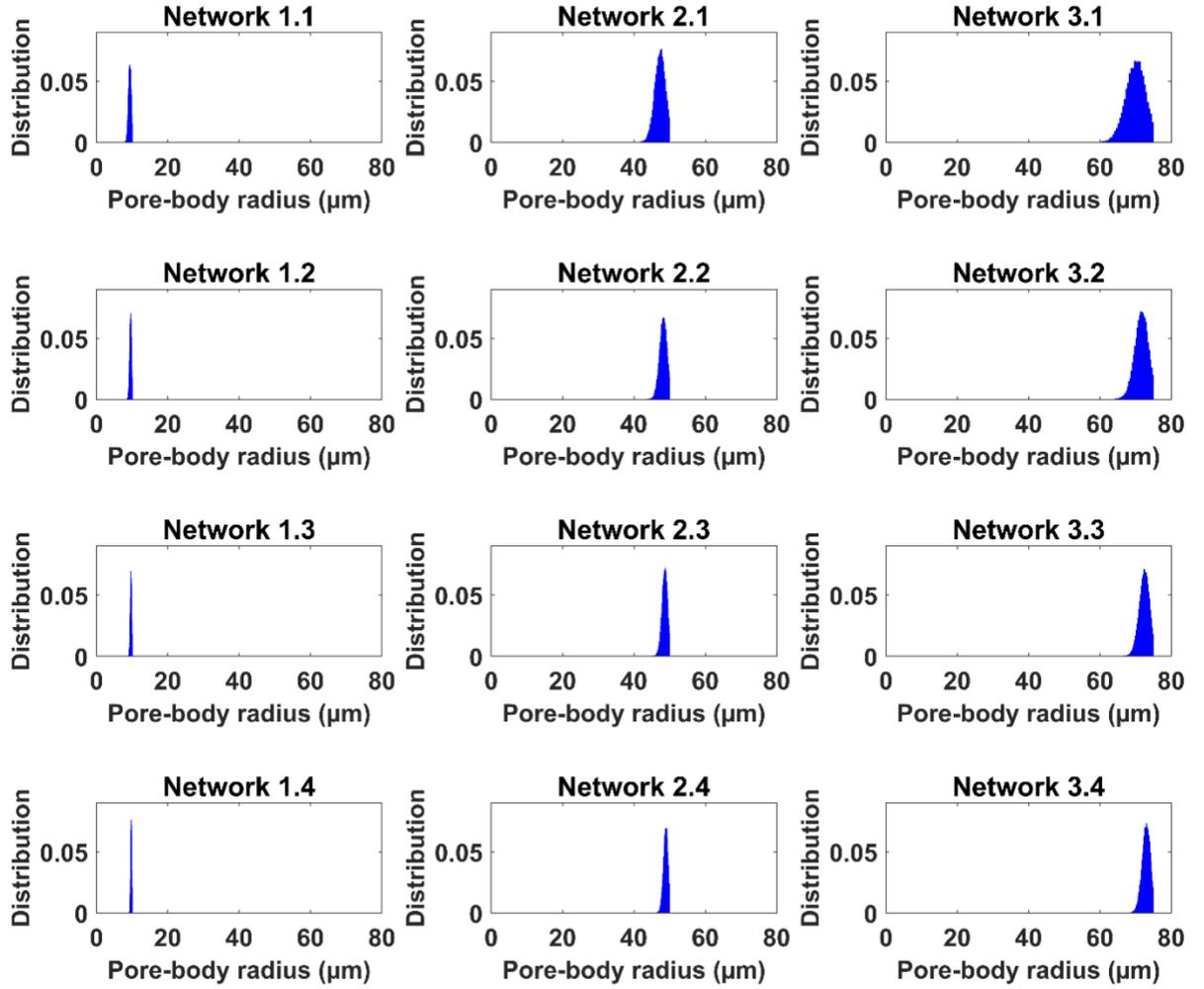

Fig. 3. Pore-body radius distributions for twelve pore networks constructed in this study.

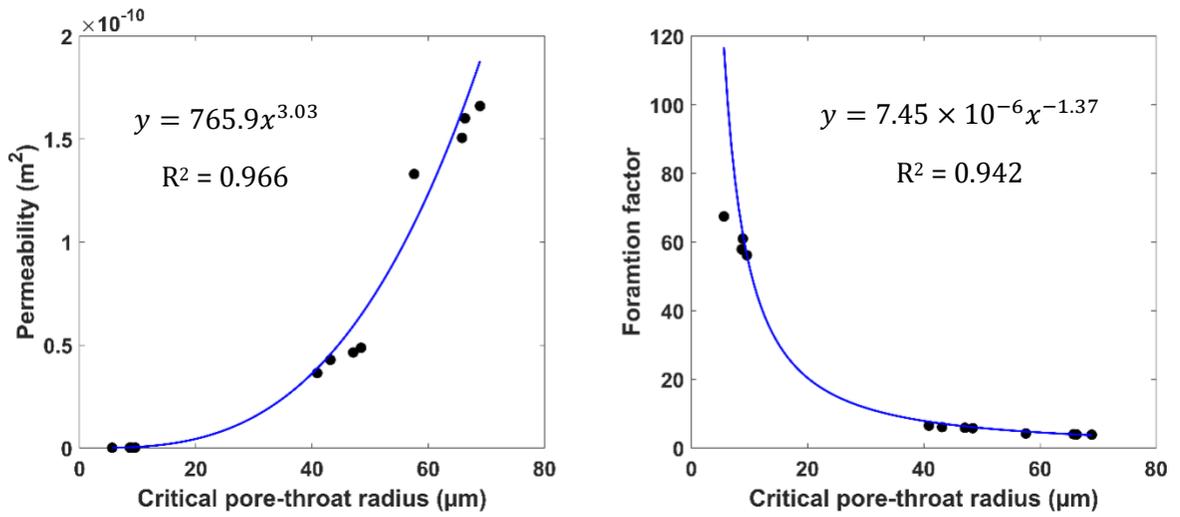

Fig. 4. Permeability (left) and formation factor (right) versus critical pore-throat radius. Blue lines are the best power-law function fitted to the simulations results. Both permeability and formation factor simulations are from the smallest network with $L = 1130$ $\mu m$. The critical pore-throat radius was determined from the mode of the pore-throat radius distribution that does not vary with network size.

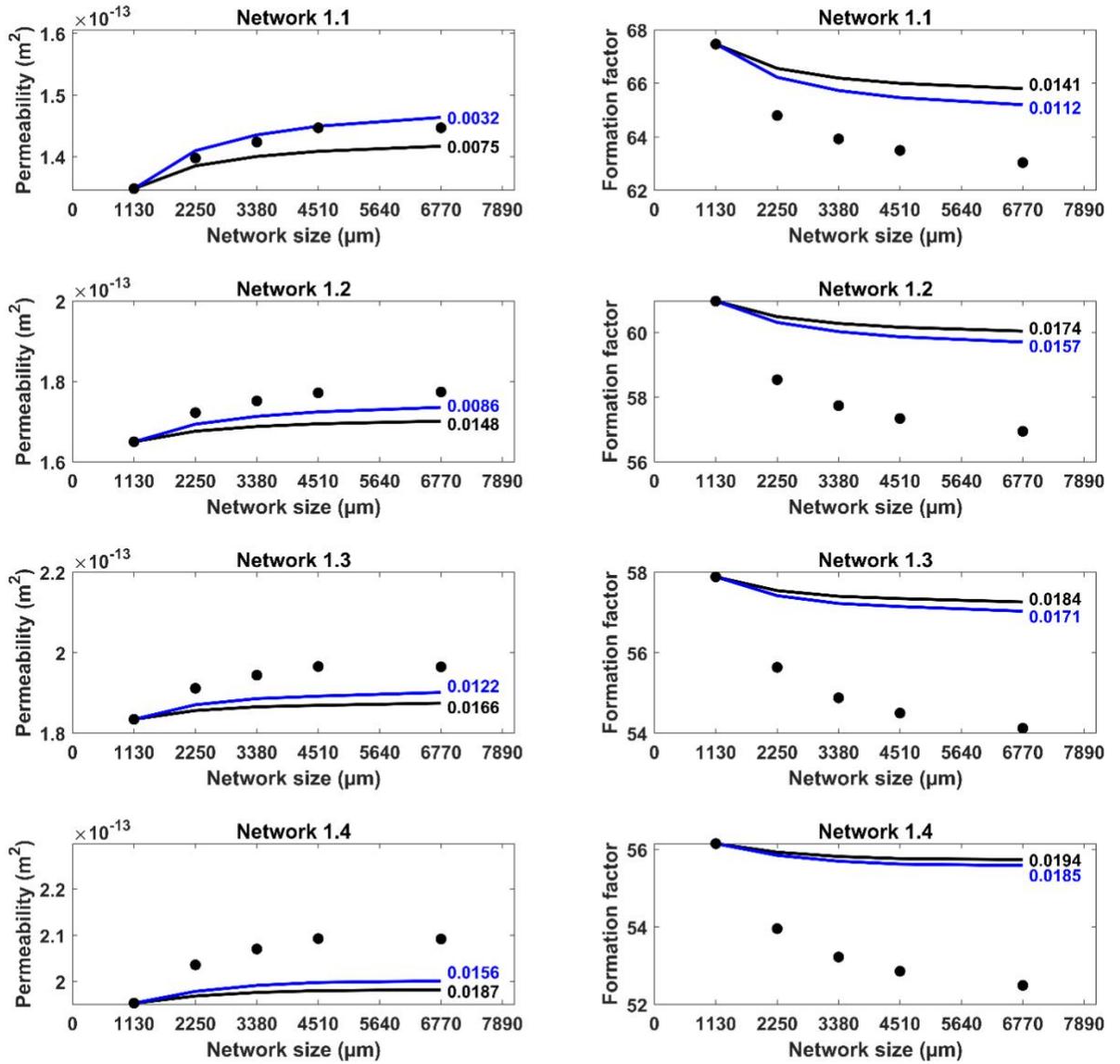

Fig. 5. Simulated and estimated permeability and formation factor for Network 1. Black filled circles indicate simulation values. Black and blue lines represent estimations by Eqs. (23) and (24) for the permeability and by Eqs. (25) and (26) for the formation factor, respectively. The calculated RMSLE value for each model is given adjacent to each line using the same color code.

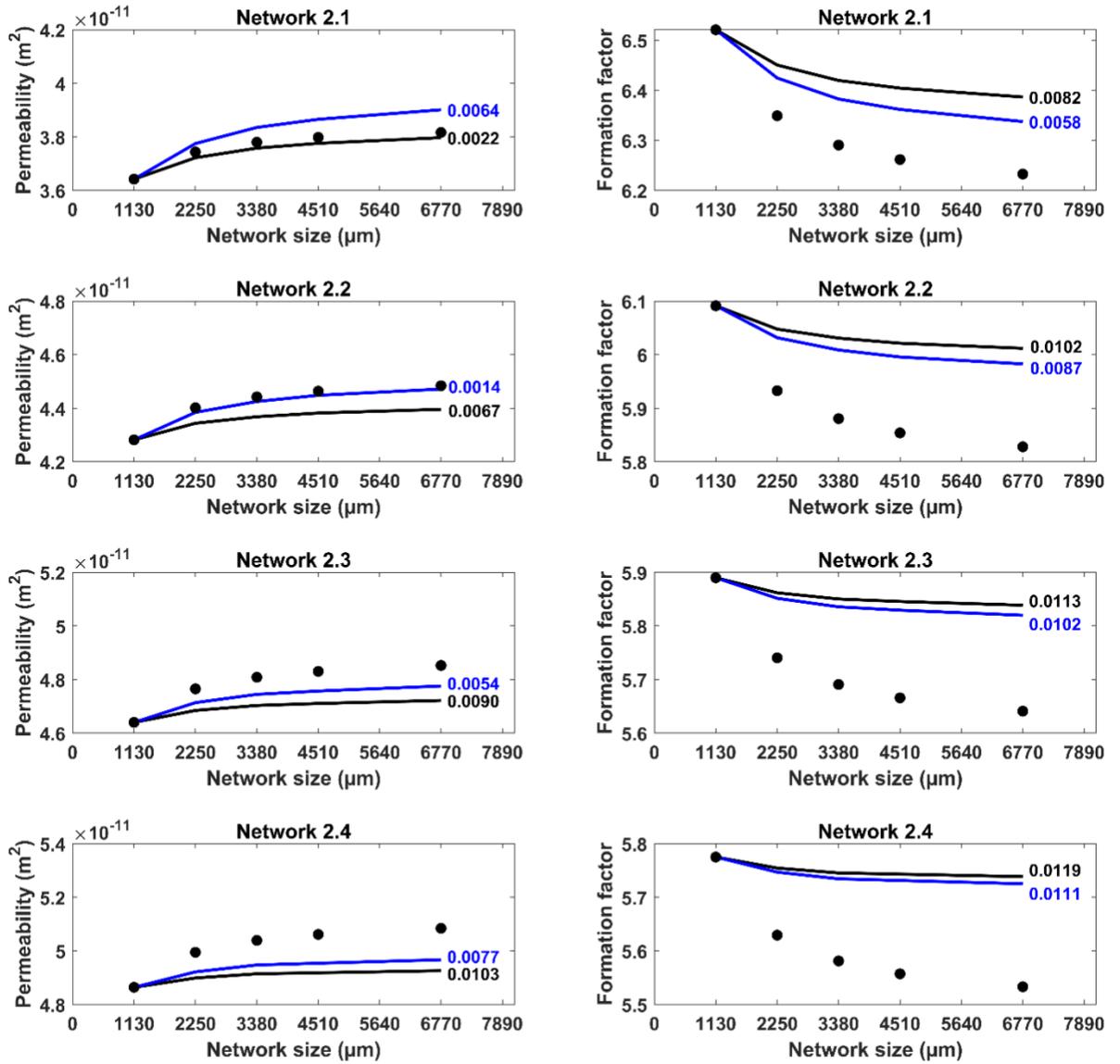

Fig. 6. Simulated and estimated permeability and formation factor for network 2. Black filled circles indicate simulation values. Black and blue lines represent estimations by Eqs. (23) and (24) for the permeability and by Eqs. (25) and (26) for the formation factor, respectively. The calculated RMSLE value for each model is given adjacent to each line using the same color code.

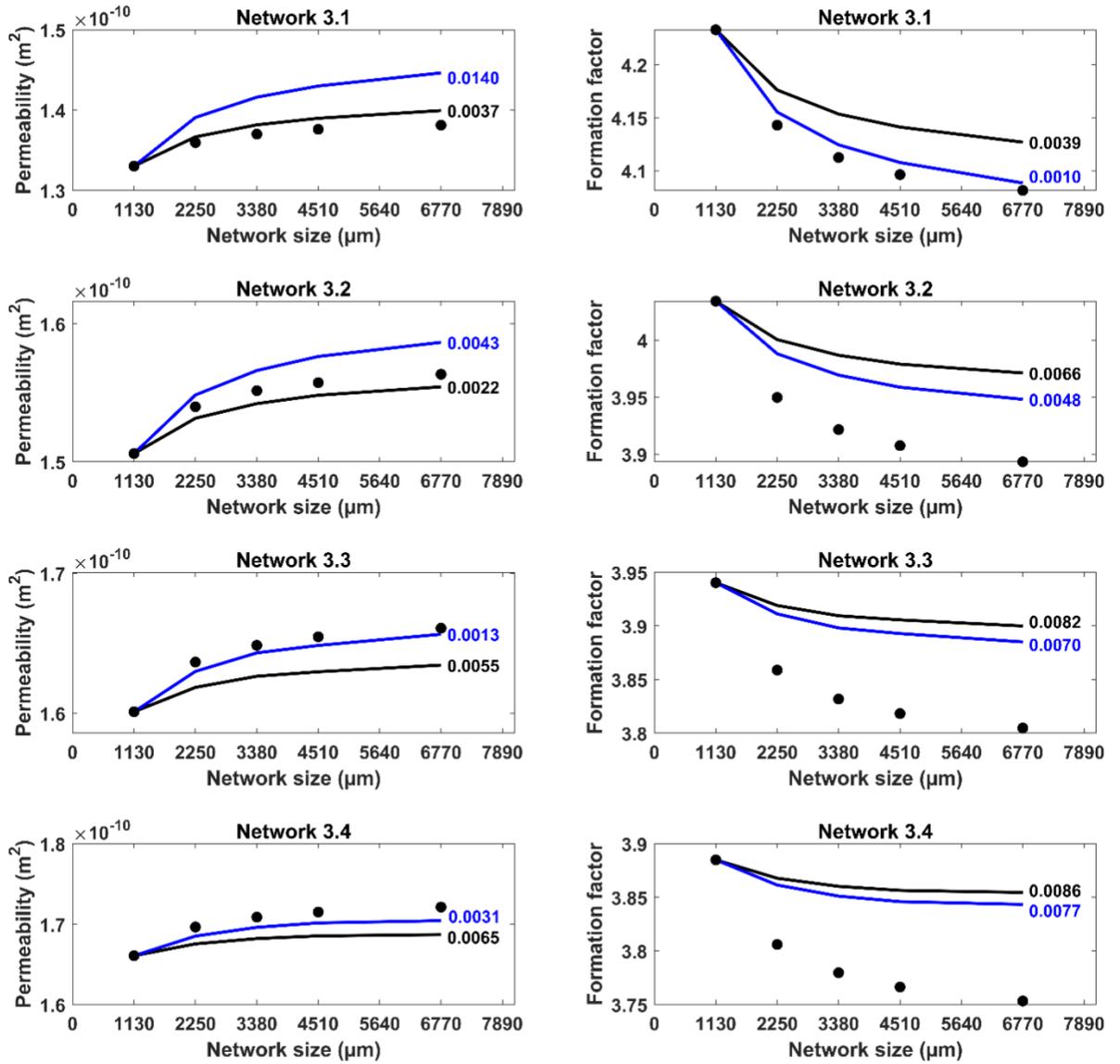

Fig. 7. Simulated and estimated permeability and formation factor for network 3. Black filled circles indicate simulation values. Black and blue lines represent estimations by Eqs. (23) and (24) for the permeability and by Eqs. (25) and (26) for the formation factor, respectively. The calculated RMSLE value for each model is given adjacent to each line using the same color code.

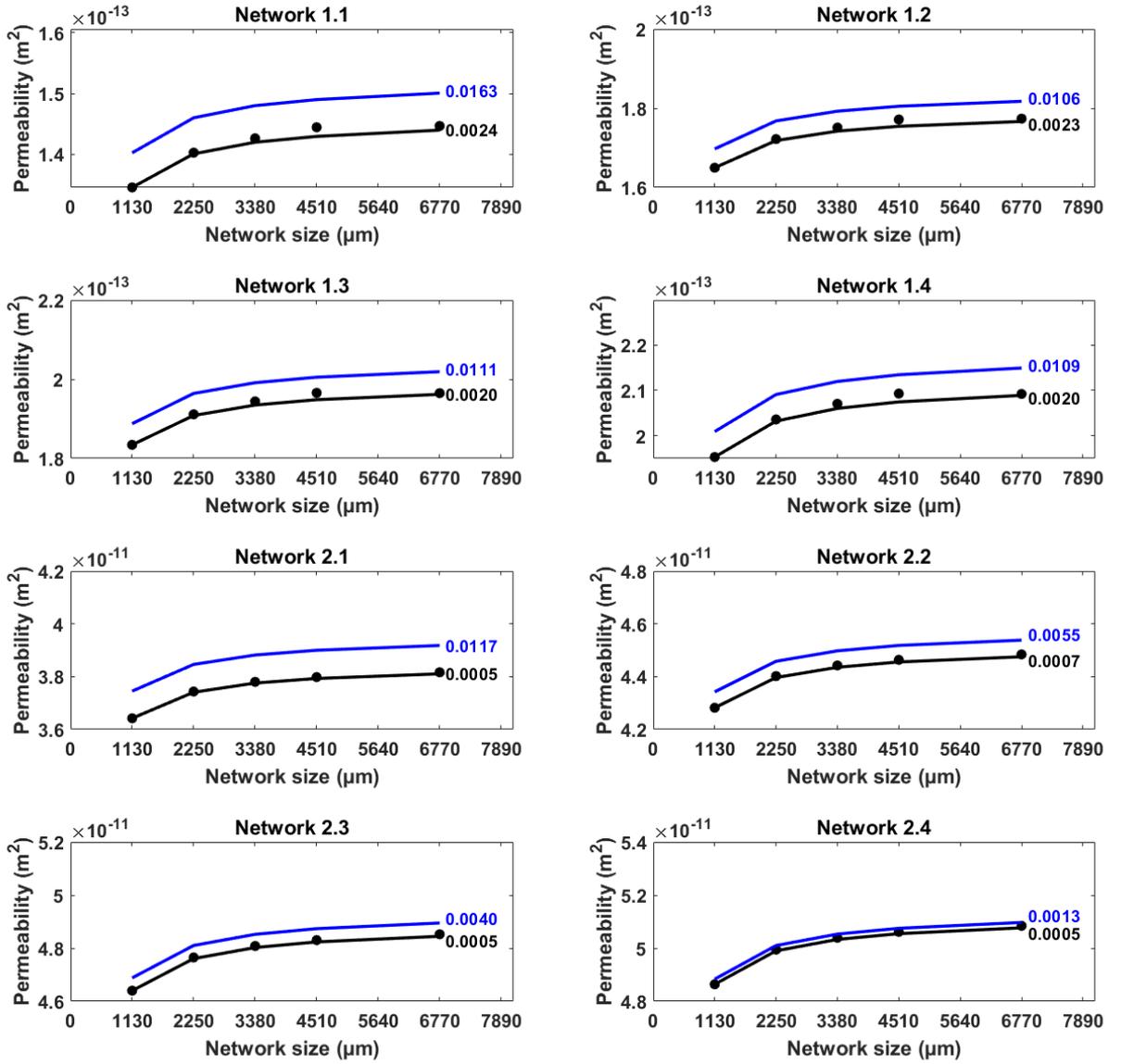

Fig. 8. Simulated and estimated permeability for the twelve pore-networks. Black filled circles indicate simulation values. Black and blue lines represent estimations by Eqs. (27) and (28), respectively. The calculated RMSLE value for each model is given adjacent to each line using the same color code.

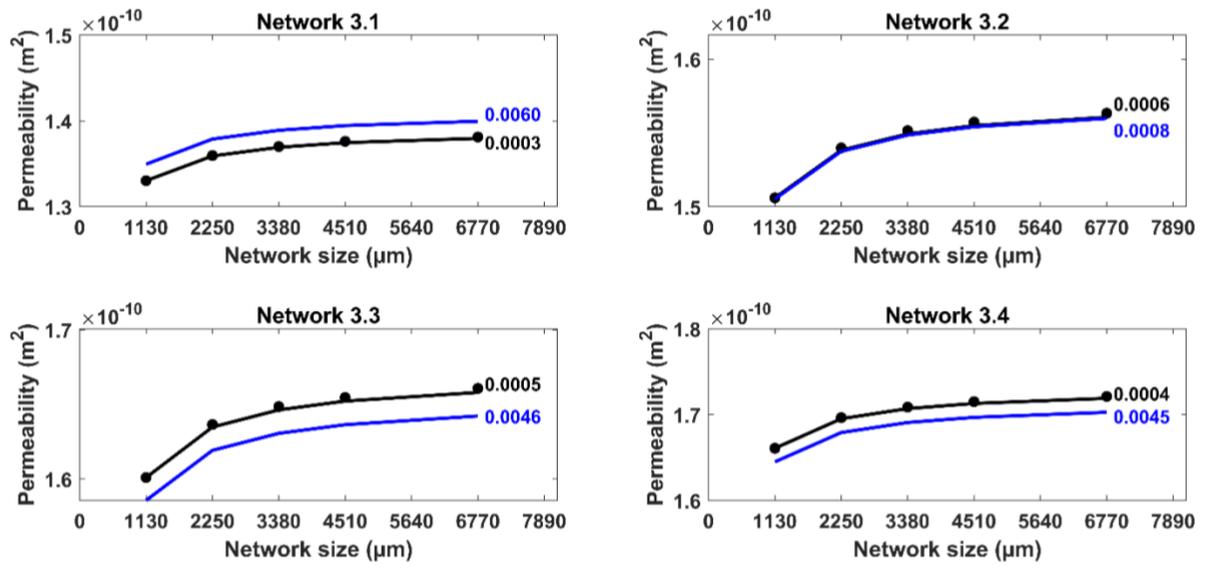

Fig. 8. (Continued)

NOTATION

| | |
|---|---|
| $A_p$ | pore cross-sectional area |
| $a_t$ | total current flow |
| $A_t$ | medium cross-sectional area |
| $A_w$ | cross-sectional area occupied by the brine |
| $C$ | constant coefficient |
| $C_{CPA}$ | Critical path analysis constant |
| $D_p$ | pore space fractal dimension |
| $F$ | formation factor |
| $f$ | volume fraction |
| $f_c$ | critical volume fraction |
| $G$ | pore shape facto |
| $g_h$ | hydraulic conductance |
| $g_e$ | electrical conductance |
| $g_{hi}$ | hydraulic conductance between the pore-throat interface and center of pore i |
| $g_{hij}$ | hydraulic conductance between two pore bodies |
| $g_{hj}$ | hydraulic conductance between the pore-throat interface and center of pore j |
| $g_{ht}$ | hydraulic conductance of pore-throat |
| $k$ | permeability |
| $k_{REV}$ | permeability at REV point[m²] |
| $L$ | network (system) size |
| $l_{b_i}$ | distance in between the center of pore body to the interface where pore body and pore throat meet |
| $l_{b_j}$ | distance in between the center of pore body to the interface where pore body and pore throat meet |
| $l_{ij}$ | distance between the centers of the two pore bodies |
| $l_{t0}$ | typical pore-throat length |
| $l_t$ | pore-throat length |
| $m$ | empirical exponent |

| Symbol | Description |
|---|---|
| $P_{inlet}$ | pressure at inlet |
| $P_{outlet}$ | pressure at outlet |
| $p$ | fraction of occupied or preset bonds |
| $p_c$ | percolation threshold |
| $q_{ij}$ | the flow rate between two pore bodies |
| $q_t$ | total flow rate |
| $R_o$ | resistivity of saturated medium |
| $r_b$ | Pore- body radius |
| $r_t$ | pore-throat radius |
| $r_{tc}$ | critical pore-throat radius |
| $r_{tmax}$ | maximum pore-throat radius |
| $r_{tmin}$ | minimum pore-throat radius |
| $R_w$ | brine resistivity |
| $V$ | pore volume |
| $V_c$ | critical volume |
| $V_s$ | sample volume |
| $V_t$ | total volume of pores |
| $x_{est}$ | estimated value |
| $x_{sim}$ | simulated value |
| $\alpha$ | CPA exponent for permeability |
| $\beta$ | CPA exponent for formation factor |
| $\gamma$ | Weibull distributions shape parameter |
| $\Delta P_{ij}$ | pressure difference between two pore bodies |
| $\Delta V$ | voltage drop |
| $\delta$ | Weibull distributions shape parameter |
| $\zeta$ | aspect ratio for pore-body radius generation |
| $\mu$ | fluid viscosity |
| $\nu$ | correlation length scaling exponent |
| $\chi$ | correlation length |
| $\chi_0$ | typical bond length |